\documentclass[%
 reprint,
 amsmath,amssymb,
 aps,nofootinbib
]{revtex4-2}

\usepackage{graphicx}
\usepackage{dcolumn}
\usepackage{bm}
\usepackage{mathtools}
\usepackage{physics}
\usepackage{amsmath}

\newcommand{\mum}{$\rm\mu m$ }

\newcommand{\Er}{\mbox{ $\rm E_r$}}

\newcommand{\Rb}{\ensuremath{^{87}}Rb}

\newcommand{\eq}[2]{\begin{equation}#1\label{#2}\end{equation}}

\newcommand{\Hop}{\hat{\mathrm{H}}}

\newcommand{\ie}{\textit{i.e. }}

\begin{document}

\preprint{APS/123-QED}

%\title{Direct reconstruction of an optical lattice's band structure}
\title{Direct reconstruction of the band structure of a 1D optical lattice with thermal atoms}

\author{Arnaud Courvoisier$^*$}
\author{Amruta Gadge$^*$}
\author{Nir Davidson$^\dagger$}
\affiliation{Department of Physics of Complex Systems, Weizmann Institute of Science, Rehovot 761001, Israel}

\date{\today}
% \begin{abstract}
% We report on the observation of sub-recoil momentum classes while performing a Kapitza-Dirac experiment with thermal clouds of \Rb\ atoms. With the support of numerical calculations, we explain their origin and show how they can be used to directly reconstruct the full band-structure of a standing wave potential. While this can serve as a precise lattice depth calibration tool, we additionally propose a method to estimate the potential depth in a single-shot manner.   
% \end{abstract}

\begin{abstract}
We report on a simple method to reconstruct the band structure of a 1D optical lattice using a thermal cloud with a momentum spread of about two-photon recoils. We image the momentum distribution of a thermal cloud exposed to a standing wave potential using time-of-flight absorption images and observe unique features. With the support of numerical calculations, we explain their appearance and show how they can be used to reconstruct the full band-structure directly. While this can serve as a precise lattice depth calibration tool, we additionally propose a method to estimate the lattice depth in a single-shot manner. 
\end{abstract}

\maketitle
\def\thefootnote{*}\footnotetext{These authors contributed equally to this work.}
\def\thefootnote{$\dagger$}\footnotetext{Corresponding author : nir.davidson@weizmann.ac.il}

\section{\label{sec:introduction}Introduction}

Periodic optical potentials can be created by interfering pairs of counter-propagating light beams \cite{bloch_lattices_review}.  Since the demonstration of the interference of Bose-Einstein condensates (BEC)  trapped in a periodic potential created by an optical lattice \cite{Anderson1998}, several applications of atom-lattice systems have emerged \cite{jeff_excitation_sprectrum,probing_superfluids_lattices,gross_bloch_quantum_simulations,quantum_walk_lattices}. For instance, degenerate atomic gases loaded in optical lattices have proven to be excellent tools for the study of condensed matter and many-body physics phenomena \cite{dalibard_many_body}. The ability to control and manipulate the lattice parameters \cite{novel_optical_lattices} as well as interactions between atoms have notably made possible the observation of the superfluid to Mott insulator phase transition \cite{Greiner2002}, and of topological states of matter \cite{topological_matter_lattices}. Coherent diffraction of atomic wave-packets from standing waves have given rise to atom-interferometry applications \cite{Philips1999,Tino_2021},  and was used in the first demonstrations of bosonic amplification \cite{ketterle_superradiance,ketterle_amplification,japanese_amplification}.
For all such experiments, it is primordial to have a good understanding of the dynamics of atoms loaded in a periodic potential and to have a precise knowledge of the lattice parameters \cite{dynamics_optical_lattices_review}. 
\newline\indent
Akin to electrons in a crystal, the behavior of atoms loaded in these 'artificial crystals' can be well described within the framework of band theory \cite{kittel2018}. The dispersion relation for atoms is given by the band structure of the lattice, which depends on the depth and geometry of the lattice potential \ie on the intensity, detuning and spatial configuration of the laser beams. The most commonly used method to calibrate optical lattices is by observing the Rabi oscillations of the population in the first diffraction order when a stationary condensate is loaded into a lattice, in the so-called Kapitza-Dirac regime \cite{Denschlag2002}. Measured oscillation frequencies can then be compared to those predicted by band theory. This technique can be extended by giving a velocity to the atoms in the lattice's frame of reference, thus controlling the atoms' quasi-momentum in the lattice frame of reference. It has also been shown that by using either phase or amplitude modulation, transitions between different bands can be probed and characterized \cite{Denschlag2002}. However, these methods rely on the extremely narrow momentum distribution of condensates to accurately control the quasi-momentum of the atomic ensemble. 
% They also provide the band structure only for a single or few quasi-momenta, so the functional shape of the lattice potential needs to be assumed in advanced. OUR METHOD ALSO ASSUMES THE FUNCTIONAL SHAPE OF THE LATTICE
\newline\indent
Here, we report on a simple method to reconstruct the band structure of an optical lattice, which also serves as a precise calibration of the potential depth. In our experiment, we use thermal atomic clouds whose momentum distributions span about two photon-recoil momenta. Loading such an ensemble in a periodic potential leads to a mixture of initial quasi-momenta, $q$, spanning an entire Brillouin zone. This allows the probing of excitations between the fundamental band and higher bands for different values of $q$, in a single experimental realization. Using band theory, we numerically calculate the contribution of each band and use this insight to reconstruct the lattice band structure from experimental data. 
\newline\indent
The ability to reconstruct the entire band structure by solely scanning the lattice duration makes this method a robust yet simple tool to calibrate the depth of optical lattices precisely. 
% and to determine its spatial functional shape. 
Furthermore, we show that our approach can be used to measure the depth of an optical lattice in a single shot, with reasonable accuracy.  
\newline\indent
In section \ref{sec:exp} we describe our experimental apparatus and key observations. We then present, in section \ref{sec:band_theory}, an overview of the band theory necessary to reconstruct the lattice's band structure. In section \ref{sec:band_structure} we show results of our numerical model and use them to reconstruct lattice band structures from our experimental findings. In section \ref{sec:single_shot} we show how this tool can be used to extract lattice depths from a single shot.

\section{Experiment} \label{sec:exp}
\begin{figure}[h]
    \includegraphics[width=\columnwidth]{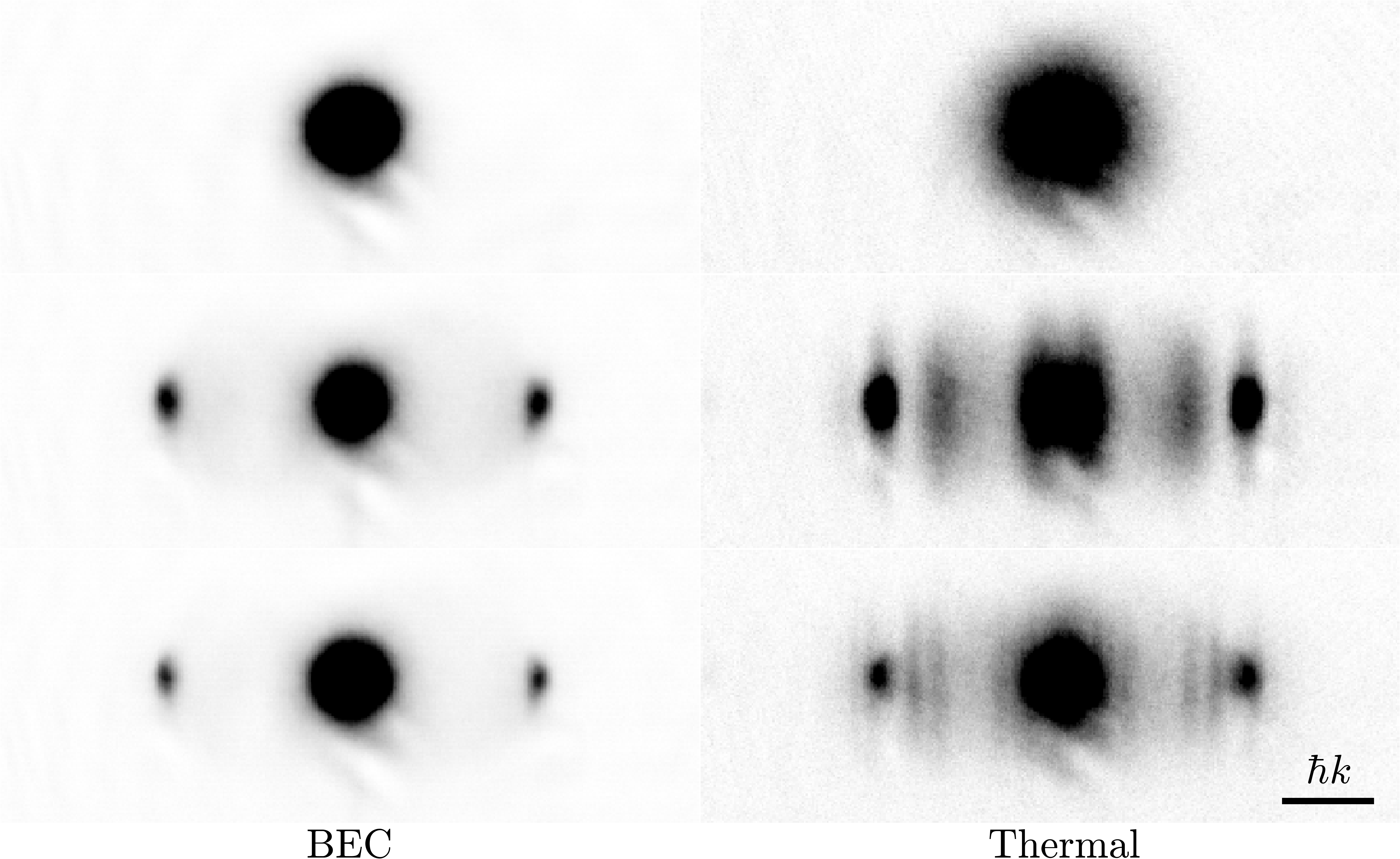}
    \caption{Absorption images after 23ms of time-of-flight for different lattice exposure times (from top to bottom, $10\mu\rm s,\ 205\mu\rm s,\ 300\mu\rm s$). The left column was measured with a pure condensate, while in the right column, the cloud was chosen to be thermal, and we observe additional features between the zeroth and first-order Kapitza-Dirac peaks. Each picture is the average of 20 experimental cycles, and the color-map was chosen to enhance faint structures.}
    \label{fig:sub_recoil_structures}
\end{figure}

\begin{figure}
\centering
    \includegraphics[width=1.005\columnwidth, trim={0cm -0.45cm -0cm -0.1cm},clip]{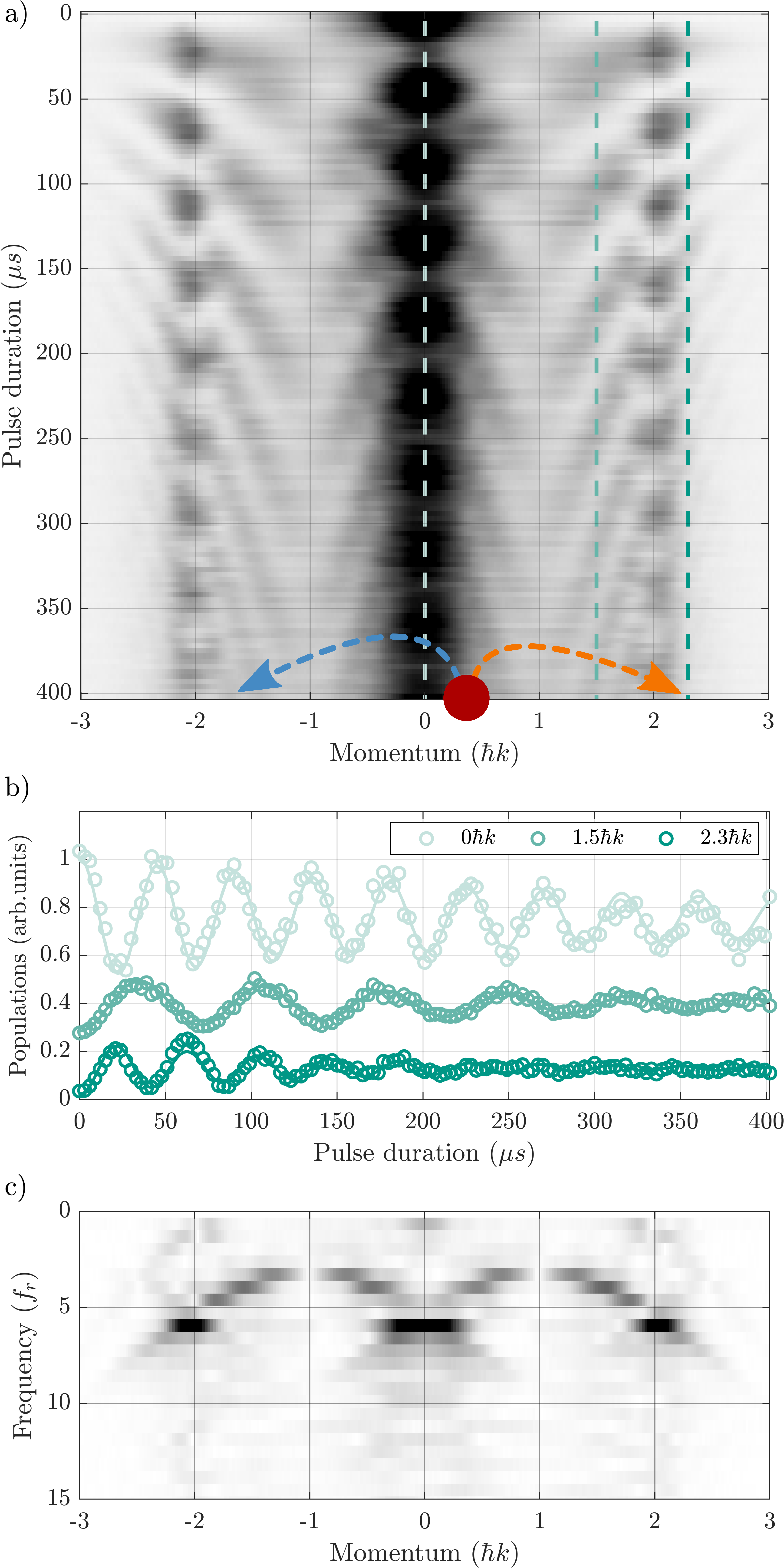}
    \caption{a) Momentum distribution as a function of the lattice pulse duration for a lattice depth of $6.5\rm E_r$. Each row is a normalized integrated density obtained from the integration of absorption images averaged 20 times. The figure is shown in an exaggerated colorbar, to reveal faint features. The overlaid dashed lines represent the values of momentum at which we measured the population oscillations shown below. The blue (orange) arrow illustrates atom transfers at a frequency corresponding to the difference between the fundamental and first (second) excited band. b) Population oscillations as a function of pulse duration for three momenta ($0\hbar k, 1.5\hbar k\text{, and } 2.3\hbar k$). Populations are shown in arbitrary units and were shifted up and down for readability. The fitted frequencies are $22\rm kHz$, $14\rm kHz$ and $24\rm kHz$, respectively. c) Fourier spectrum as a function of momentum.}
    \label{fig:population_oscillations}
\end{figure}

Our experimental system produces either  \Rb\ BECs or ultra-cold thermal gases, using an all-optical scheme. We load two overlapping far-off-resonance optical dipole traps (1.06 micron wavelength) with waists of 150\mum and $50\rm\mu m$, respectively, from a magneto-optical trap. With laser cooling followed by evaporative cooling, we obtain a nearly pure condensate with $2\times 10^5$ atoms in the $\rm F=1,m_F=-1$ state. By terminating the evaporation sequence before the cloud is condensed we obtain a thermal cloud and set the width of its momentum distribution to be $2\hbar k$ and centered around zero. The cloud is then exposed to a 1D optical lattice pulse, for durations up to 400 microseconds, during which the atoms undergo Kapitza-Dirac diffraction \cite{Denschlag2002}. We create the lattice standing wave potential by retro-reflecting a Gaussian beam with a waist of 1.05mm and a typical detuning of $-50$GHz from the $\rm2S$ to $\rm2P$ transition of \Rb. We vary the power of the lattice beams to get lattice depths, $\rm V_0$, in the $1-15\rm Er$ range, where $\Er = \hbar^2k^2/2m$, is the single-photon recoil energy. For these parameters, only the first diffraction orders $\pm2\hbar k$ are significantly populated. Throughout our parameter range, the spontaneous scattering probability per atom during a lattice pulse is less than few percent, allowing us to neglect the effects of scattering and consider only the optical dipole potential. For our thermal ensembles, both the mean-field shift due to interactions \cite{roee_rev} and the probability for momentum changing atomic collisions during a lattice pulse are sub-percent. We can therefore safely assume that the momentum manifolds [$q, q-2\hbar k, q+2\hbar k$] for different values of $q$ do not interact.    
\newline\indent
The momentum distribution of the excited atoms is measured via absorption imaging, after $23\rm ms$ of time-of-flight. Typical time-of-flight images for a pure  condensate at rest in the lattice's frame of reference ($q=0$) exposed for $10\mu\rm s,\ 205\mu\rm s, and \ 300\mu\rm s$ to a lattice depth of $6.5\rm E_r$ are presented in Fig.~\ref{fig:sub_recoil_structures}. The Kapitza-Dirac diffraction orders $\pm2\hbar k$ are known to coherently oscillate with time at a frequency \cite{Gadway2009, Grimm2000}
\eq{f_{\left|2\hbar k\right|} = \frac{1}{h}\sqrt{\left(\frac{6\pi c^2}{\sqrt{2}\omega_0^3}\right)\frac{\Gamma^2}{\Delta^2}I_0^2+\Er^2},
}{eq:bragg_frequency} 
where $I_0$ and $\Delta$  are the laser's intensity and detuning, $\Gamma$ and $\omega_0$ are the linewidth and frequency of the transition. Expression (\ref{eq:bragg_frequency}) is valid in the limit of shallow lattices \ie when $V_0\lesssim4\Er$. This technique is used extensively to calibrate the depth of optical lattices, as measurements of $f_{\left|2\hbar k\right|}$ yield accurate estimations of the field intensity experienced by the atoms.     
\newline\indent
We then repeat the experiment with a thermal cloud and observe rich features in the resulting momentum distributions, as seen in Fig.~ \ref{fig:sub_recoil_structures}. To study the time dynamic of each momentum state, we integrate the time-of-flight images along the vertical axis, which yields the momentum distribution of the cloud as a function of the lattice pulse duration, as shown in Fig.~\ref{fig:population_oscillations}.a. The top-most row in this figure indicates the initial momentum distribution of the cloud before the lattice pulse is applied, it is seen to span from about $-\hbar k$ to $+\hbar k$.  Each column presents the population dynamics corresponding to a different momentum class.
Figure ~\ref{fig:population_oscillations}.b  shows three vertical cross sections of Fig.~\ref{fig:population_oscillations}.a, depicting oscillations in the populations of  three different quasi-momenta, $q=0,1.5k \text{ and } 2.3k$, which oscillate at different frequencies. A Fourier spectrum for each momentum class is  shown in Fig.~\ref{fig:population_oscillations}.c, revealing a dominant oscillation frequency for each quasi-momentum.

In the following sections, we present a theoretical model to explain these oscillations and show how we can use them to reconstruct the lattice's band structure in the first Brillouin zone.
\medbreak

\section{Band theory for atoms loaded in a periodic potential}  \label{sec:band_theory}
Right after releasing our cloud from the optical dipole trap, we abruptly turn on the standing wave for a variable duration, before turning it off suddenly. In the sudden approximation, the initial state of the free atoms is simply projected onto the basis of the eigen-vectors of the lattice, when the latter is turned on. The different components of the new state acquire different phases while the lattice is on, and the resulting state is subsequently projected onto the plane-wave basis, when the lattice is finally turned off. The time-evolution of the different velocity classes can be derived following the derivations of \cite{Dalibard2013,Denschlag2002}.\medbreak\noindent
Let us consider a standing wave represented by a periodic potential 
\eq{V(x)=V_0\sin^2(kx),}{}
with \cite{Grimm2000},
\eq{V_0 = \frac{6\pi c^2}{\omega_0^3}\frac{\Gamma}{\Delta}I_0.}{}
According to the Bloch theorem, the eigen-states of the Hamiltonian 
\eq{\Hop = \frac{\hat{p}^2}{2m}+V_0\sin^2(k\hat{x})}{eq:hamiltonian}
can be written under the form
\eq{\psi_q(x) = e^{iqx}u_q(x),}{}
\noindent where $q$ is the quasi-momentum. The $u_q$ function is periodic in the position $x$ and as such can be decomposed in a Fourier series, and written as
\eq{u_q(x) = \sum_{j\in\mathbb{Z}}C_j^{\hspace{1pt}q}e^{2i\pi jx/a},}{}
giving
\eq{\psi_q(x) = \sum_{j\in\mathbb{Z}}C_j^{\hspace{1pt}q}e^{ix(q+2\pi j/a)}.}{eq:bloch_vector}
Using equations (\refeq{eq:hamiltonian}) and (\refeq{eq:bloch_vector}), one gets that the Fourier decomposition factors must obey
\eq{\left[\left(2j+\frac{q}{k}\right)^2+\frac{V_0}{2\rm E_r}\right]C_j^{\hspace{1pt}q}-\frac{V_0}{4\rm E_r}\left(C_{j-1}^{\hspace{1pt}q}+C_{j+1}^{\hspace{1pt}q}\right)=\frac{E}{\rm E_r}C_j^{\hspace{1pt}q}.}{eq:mathieu}
We denote by $E_n^{\hspace{1pt}q}$ the eigen-values associated to band $n$ and to quasi-momentum $q$. The corresponding Fourier decomposition factors are denoted by $C_{n,j}^{\hspace{1pt}q}$ and we write the Bloch vector associated to band $n$ and quasi-momentum $q$ as 
\eq{\ket{\psi_{n,q}(x)} = \sum_{j\in\mathbb{Z}}C_{n,j}^{\hspace{1pt}q}\ket{\phi_{q+2j k}},}{}
where $\ket{\phi_{q+2j k}}$ represents the plane-wave state with momentum $\hbar q+2j\hbar k$. Let us consider atoms with an initial momentum $\hbar q_i$ in the standing wave reference frame. We denote the initial state by $\ket{\phi_{q_i}}$. At time $t = t_i$, we abruptly turn on the lattice and project the initial state onto the lattice eigenstates such that 
\eq{\ket{\psi(t_i)} = \sum_{n=0}^\infty{\ket{\psi_{n,q_i}}\braket{\psi_{n,q_i}}{\phi_{q_i}}}=\sum_{n=0}^\infty{C_{n,0}^{\hspace{1pt}q_i*}\ket{\psi_{n,q_i}}}.}{}
Each eigen-state then proceeds to acquire a phase proportional to its energy, such that at a subsequent time $t\geq t_i$, the wave function becomes
\eq{\ket{\psi(t)} = \sum_{n=0}^\infty{C_{n,0}^{\hspace{1pt}q_i*}\ket{\psi_{n,q_i}}e^{-i E_n^{\hspace{1pt}q_i}(t-t_i)/\hbar}}.}{}
Turning off the lattice abruptly at time $t = t_f$ then amounts to projecting back onto the plane-wave basis and one gets that 
\eq{\ket{\psi(t_f)} = \sum_{j\in\mathbb{Z}}\left\{\sum_{n=0}^\infty C_{n,0}^{\hspace{1pt}q_i*}C_{n,j}^{\hspace{1pt}q_i}e^{-iE_n^{\hspace{1pt}q_i}(t_f-t_i)/\hbar}\right\}\hspace{-2pt}\ket{\phi_{q_i+2j k}}\hspace{-2pt}.}{eq:band_contribution}
When a pure BEC is loaded into the lattice, the initial state is well approximated by $\ket{\phi_{q_i=0}}$. In this case, because of their opposite parity, the odd bands do not take part in the time evolution \cite{Gadway2009}. For weak lattices, this results in oscillations in the population of the first Kapitza-Dirac orders, at a frequency corresponding to the energy gap between the fundamental and second-excited bands. However, these symmetry considerations do not hold for non-zero initial quasi-momenta, and any pair of bands can dominate the temporal dynamic.
\newline\indent
When limiting  the analysis to the fundamental and the first two excited bands, valid for weak lattices, we find that the probability $P^{\hspace{1pt}q_i+2k}_{q_i}(t)$ for atoms originally in $\ket{\phi_{q_i}}$ to be in state $\ket{\phi_{q_i+2k}}$ after an exposure time $t$, reads
\eq{
\begin{split}
P^{\hspace{1pt}q_i+2k}_{q_i}(t) &= \sum_{n=0}^2\abs{C_{n,0}^{\hspace{1pt}q_i*}C_{n,1}^{\hspace{1pt}q_i}}^2\\
&+ 2A_{0,1}^{1,q_i}\cos\left[\frac{E_0^{\hspace{1pt}q_i}-E_1^{\hspace{1pt}q_i}}{\hbar}(t_f-t_i)+\varphi_{0,1}^{1,q_i}\right]\\
&+ 2A_{0,2}^{1,q_i}\cos\left[\frac{E_0^{\hspace{1pt}q_i}-E_2^{\hspace{1pt}q_i}}{\hbar}(t_f-t_i)+\varphi_{0,2}^{1,q_i}\right]\\
&+ 2A_{1,2}^{1,q_i}\cos\left[\frac{E_1^{\hspace{1pt}q_i}-E_2^{\hspace{1pt}q_i}}{\hbar}(t_f-t_i)+\varphi_{1,2}^{1,q_i}\right],
\end{split}
}{eq:weak_lattices_dec}
where 
\eq{A_{n,m}^{j,q_i} = \abs{C_{n,0}^{\hspace{1pt}q_i*}C_{n,j}^{\hspace{1pt}q_i}}\times\abs{C_{m,0}^{\hspace{1pt}q_i}C_{m,j}^{\hspace{1pt}q_i*}},}{}
and $\varphi_{n,m}^{\hspace{1pt}j,q_i}$ is the phase originating from the arguments of complex prefactors, such that
\eq{\varphi_{n,m}^{\hspace{1pt}j,q_i} = \arg\left[C_{n,0}^{\hspace{1pt}q_i*}C_{n,j}^{\hspace{1pt}q_i}\right]-\arg\left[C_{m,0}^{\hspace{1pt}q_i}C_{m,j}^{\hspace{1pt}q_i*}\right].}{}
We will see in the next section that, in practice, only one of the oscillating terms contributes significantly to the time dynamic for weak lattices. This allows us to separate the influence of each band, thereby fully reconstructing the relevant band structure.  

\section{Reconstruction of the optical lattice's band structure} \label{sec:band_structure}
\begin{figure}[h]
    \centering
    \includegraphics[width = \columnwidth, trim={0cm 0cm 0cm 0cm},clip]{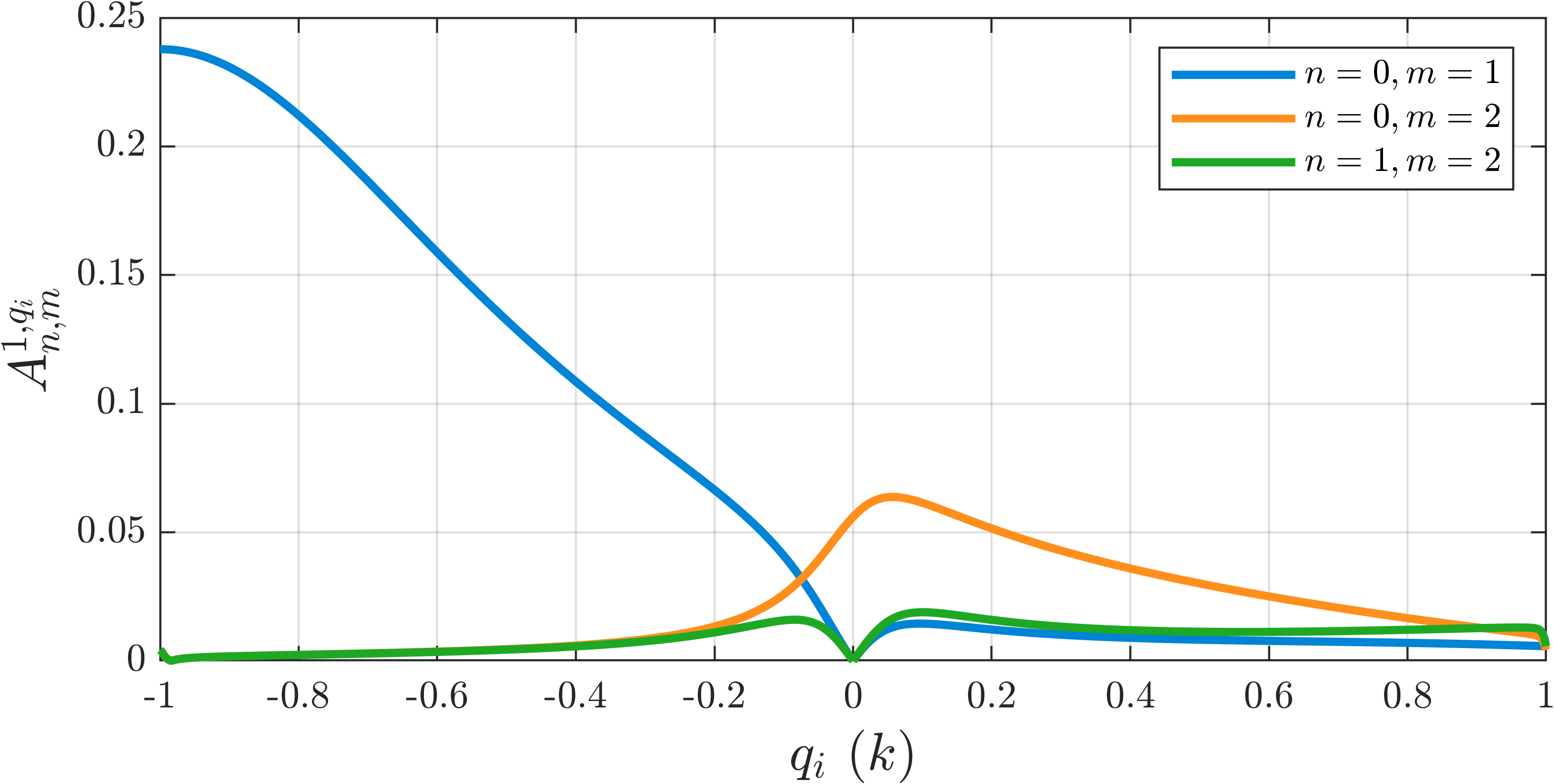}
    \caption{Numerical calculation of the band decomposition coefficients when an atom is transferred from $q_i$ to $q_i+2k$, as a function of the initial quasi-momentum $q_i$, for $V_0 = 5\rm E_r$. We only plot the three terms of equation (\ref{eq:weak_lattices_dec}) as, for weak lattices, the contribution of higher bands is significantly suppressed. In the case of atoms transferred from $q_i$ to $q_i-2k$, the curves would simply be flipped around $q_i = 0$.}
    \label{fig:band_contribution}
\end{figure}

\begin{figure}
    \centering
    \includegraphics[width = \columnwidth]{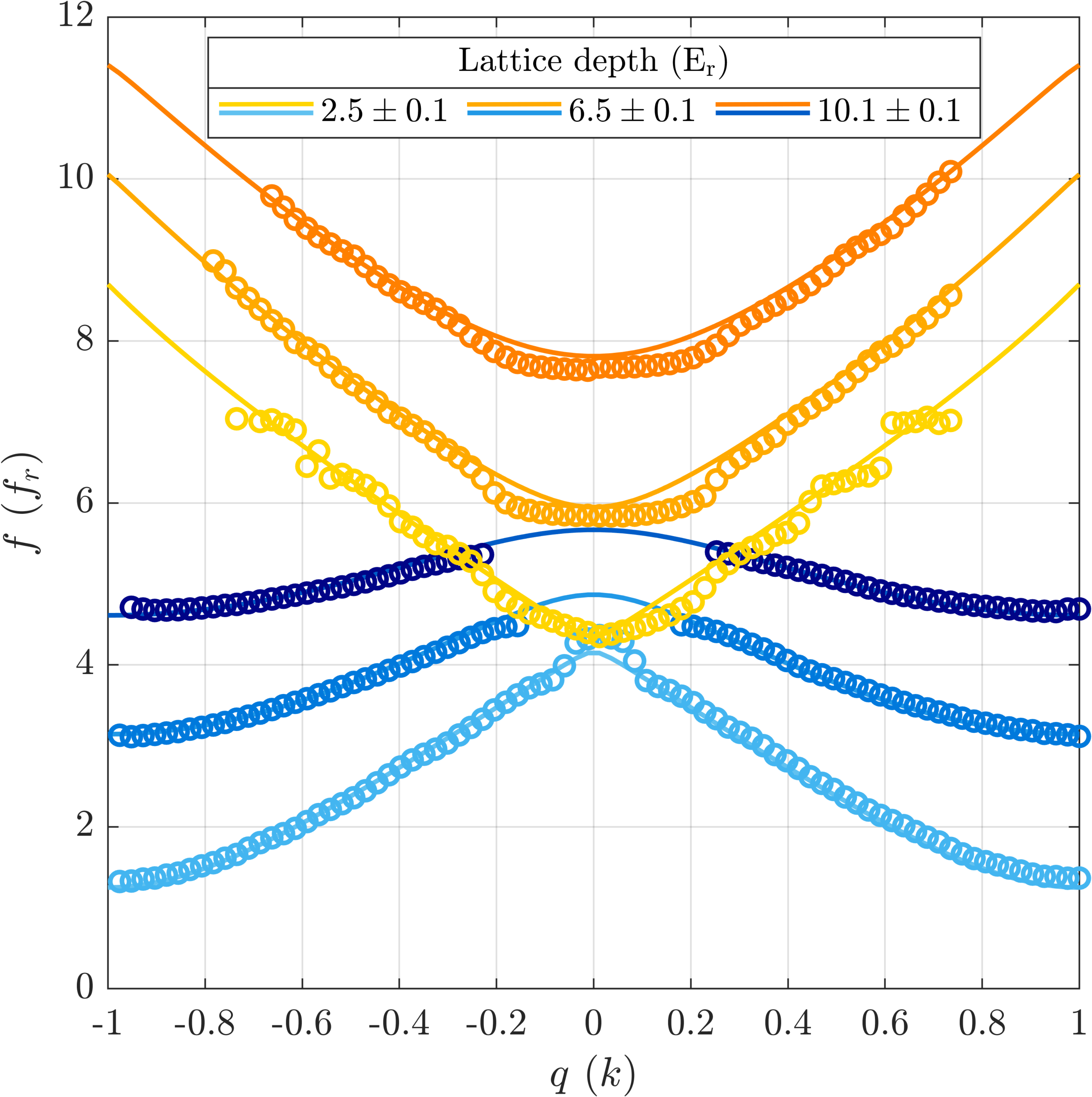}
    \caption{Frequency difference, $f$, between the fundamental and first excited bands (blue) as well as the frequency difference between the fundamental and second excited band (orange) for various lattice depths. Solid lines are a fit to numerical band-structure calculations, using lattice depth as the sole fit parameter. The fitted lattice depths are $V_0 = 2.5\pm0.1\rm E_r$, $6.5\pm0.1\rm E_r$ and $10.1\pm0.1\rm E_r$.}
    \label{fig:bands}
\end{figure}

The interesting time dynamic observed in our experiment originates from the fact that we initially populate non-zero quasi-momenta and observe the contribution of both the first and second bands separately. Equation (\ref{eq:band_contribution}), indicates that the contribution of the $n$-th band to the time evolution of the population in the plane-wave state $\ket{\phi_{q_i+2jk}}$ is weighted by the factor $C_{n,0}^{\hspace{1pt}q_i*}C_{n,j}^{\hspace{1pt}q_i}$. Solving equation (\ref{eq:mathieu}) numerically for a lattice depth of $V_0 = 5\rm E_r$, we compute the value of the contribution of the fundamental and the $n$-th excited band, $A_{n,m}^{1,q_i}$, as a function of the initial quasi-momentum $q_i$. The corresponding simulation results are shown in Fig.~\ref{fig:band_contribution}.
\newline\indent
For initial quasi-momenta, $q_i\gtrsim -0.1k$, the time evolution of atoms getting transferred to $q_i+2k$ is mostly dictated by the fundamental and the second bands, as shown in Fig.~\ref{fig:band_contribution}. Conversely, only the first band contributes significantly to the dynamic of atoms starting with quasi-momenta $q_i\lesssim -0.1k$ and gaining $2k$. The first band does not contribute to the evolution for $q_i=0$ due to the symmetry argument previously discussed. Alternatively, the band contributions for transferring atoms from $q_i$ to $q_i-2k$ have the opposite behavior to that seen in Fig.~\ref{fig:band_contribution}. In Fig.~\ref{fig:population_oscillations}.a, we show in a cartoon-like manner with an orange (blue) arrow how atoms with initial quasi-momenta comprised between $0$ and $+k$ will mainly gain (lose) a momentum $2\hbar k$ with an oscillation frequency corresponding to the difference between the fundamental and the second (first) excited bands.
\newline\indent
These results can be used to understand the experimental findings shown in Fig.~\ref{fig:population_oscillations}. For example, atoms with momenta between $\pm2\hbar k$ and $\pm3\hbar k$ will mainly oscillate between the fundamental and the second band, whereas atoms with momenta between $\pm1\hbar k$ and $\pm2\hbar k$ will mainly make transitions to the first band. Using this understanding, we map the observed data onto the first Brillouin zone \ie $q_i$ lying between $\pm1 k$. Once we reduce the data to that region, we fit the oscillations in the population of $q_i$ with an exponentially decaying sine function and use the fitted frequencies to reconstruct the lattice's band-structure, as presented in figure \ref{fig:bands}. Fitting our data to theory predictions provides a precise measurement of the lattice depth based on numerous data points, for the same experimental cost as the standard method of fitting only the $\pm2\hbar k$ populations.
\newline\indent
For lattice depths in the $V_0\gtrsim4\rm E_r$ regime, the fundamental band is relatively flat with respect to the excited bands, meaning that what we measure can be approximated to the bands themselves. For instance, using numerical calculations, we estimate that the second excited band differs by less than $3\%$ from what we measure, at $V_0 = 10\rm E_r$. One can clearly observe the expected flattening of the bands as the lattice depth is increased. 
\section{Single-shot lattice depth calibration}\label{sec:single_shot} 
\begin{figure}[h]
    \centering
    \includegraphics[width = 1\columnwidth]{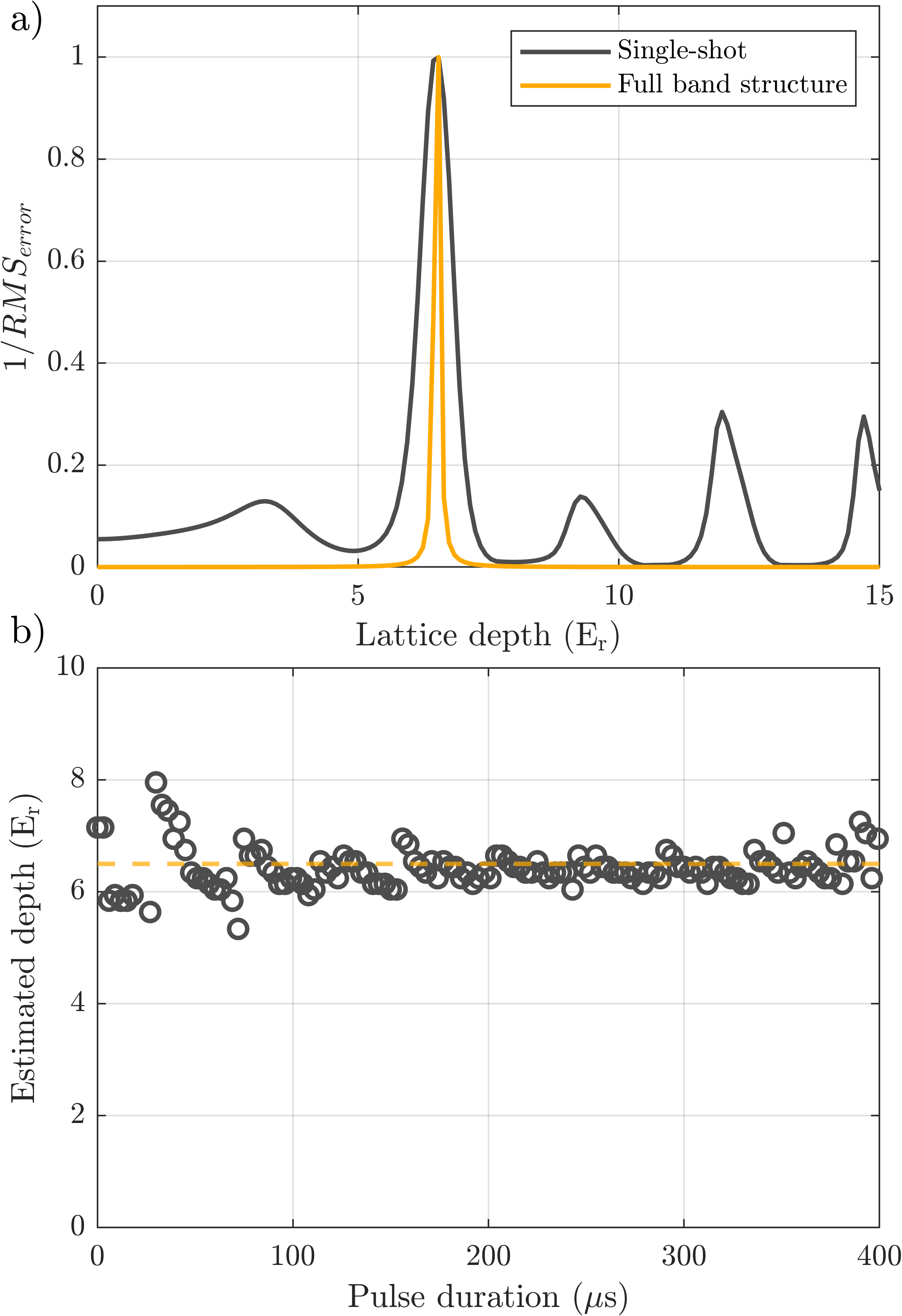}\vspace{4pt}
    \caption{a) In gray (orange) is shown the inverse of the RMS difference between a single-picture momentum distribution (full band structure) and theoretically predicted momentum distribution for different lattice depths, showing a sharp response around the estimated depth. In the single-shot case, the lattice exposure time was chosen to be $177\rm\mu s$. The full-width-at-half-maximum of the single-shot (full band structure) peak is $0.9\rm E_r$ ($0.15\rm E_r$). b) Data points show the single-shot lattice depth estimation as a function of the lattice exposure time. The horizontal dashed line corresponds to the lattice depth as measured by the full band structure method.}
    \label{fig:single_shot}
\end{figure}

Using equation (\ref{eq:band_contribution}) and the uniqueness of the $C_{n,j}^{\hspace{1pt}q_i}$ coefficients for a given potential depth, one can see that the momentum distribution after a given time in the lattice uniquely depends on the lattice depth. Therefore, using a single thermal-cloud picture after a specific lattice duration, such as those shown in Fig.~\ref{fig:sub_recoil_structures}, one can uniquely determine the lattice depth. We have developed a program \cite{github} that takes as an input a time-of-flight picture and the corresponding lattice exposure time and outputs an estimation of the lattice depth. The program solves equation (\ref{eq:band_contribution}) for the specified lattice exposure time and generates momentum distributions for various lattice depths. In order to simulate a finite imaging resolution, we apply a Gaussian blur on the generated momentum distributions, which are then compared to the experimental one. The program outputs the potential depth that minimizes the RMS-error between simulated and experimental data as an estimate of the experimental lattice depth. This is exemplified in Fig.~\ref{fig:single_shot}.a, for a lattice depth of $V_0 = 6.5\rm E_r$ and a lattice exposure time of $177\rm\mu s$.  
\newline\indent
For exposure times that are short with respect to the lattice oscillation period, the difference between momentum distributions originating from different lattice depths is too small for the estimate to be accurate. For very long exposure times, the decay of the oscillations diminishes the contrast of the data and hinders our ability to discern different lattice depths. We therefore systematically study which exposure times yield the best results and find that for lattices with depths in the $1-15\rm E_r$ range, exposure times between $100\rm\mu s$ and $350\rm\mu s$ yield accurate results, as demonstrated in Fig.\ref{fig:single_shot}.b. The uncertainty of the method stems both from the spread of the estimates shown in Fig.~\ref{fig:single_shot}.b, which amounts to $0.2\rm E_r$ for times comprised between $100\rm\mu s$ and $350\rm\mu s$, and from the width of the single-shot peak in Fig.~\ref{fig:single_shot}.a.  In the case shown in Fig.~\ref{fig:single_shot}.a, the measured lattice depth is therefore $V_0 = 6.5\pm0.4\rm E_r$, which is in accordance with the value we measured using the full band structure calibration. 
\medbreak
\vspace{0pt}
\vspace{0pt}
\section{Conclusion}\vspace{0pt}In conclusion, we reported on the presence of unique features observed in our absorption images after long times-of-flight, which arise due to the wide initial quasi-momentum distribution of our thermal cloud when it is loaded in a 1D optical lattice. Using a numerical model based on band theory, we interpreted these features and extracted the band structure of the lattice. This method proves to be simple as it only requires scanning the lattice pulse duration and does not require a Bose-Einstein condensate. The broadly used lattice calibration technique of measuring oscillation frequencies of populations at $\pm2\hbar k$ only yields a single point in the band diagram of the lattice, whereas performing the same experiment with a mixture of initial quasi-momenta yields the entire band structure of the lattice. Our experiment can therefore be used as a simple and precise calibration technique for optical lattices even in cases where their spatial potential is not known in advance. Furthermore, effects of atomic interactions and collisions during the lattice exposure, that are often significant for Bose-Einstein condensates \cite{bloch_lattices_review, dalibard_many_body},  can be safely neglected for our thermal atoms due to their much lower densities. Finally, through a detailed understanding of the time dynamic of atoms in an optical lattice, we presented a new single-shot lattice depth calibration method and shared the required image analysis program.  
\section*{Acknowledgements}
This work was supported by the Israel Science Foundation grant 1314/19.  The authors acknowledge fruitful discussions with Boaz Raz and Gavriel Fleurov. 
\bibliography{biblio}

\end{document}